\documentclass[12pt]{iopart}
\usepackage{graphicx}
\usepackage{bm}

\begin{document}
\title[Zitterbewegung of electrons in semiconductors]{Zitterbewegung of nearly-free and tightly bound electrons in semiconductors}
\date{\today}

\author{T M Rusin $^1$ and W Zawadzki$^2$}
\address{$^1$PTK Centertel Sp. z o.o., ul. Skierniewicka 10A, 01-230 Warsaw, Poland}
\address{$^2$ Institute of Physics, Polish Academy of Sciences,Al. Lotnik\'ow 32/46, 02-688 Warsaw, Poland}
\ead{Tomasz.Rusin@centertel.pl}

\begin{abstract}
 We show theoretically that nonrelativistic nearly-free electrons in
solids should experience a trembling motion
 (Zitterbewegung, ZB) in absence of external fields, similarly to
relativistic electrons in vacuum.
 The Zitterbewegung is directly related to the influence of periodic
potential on the free electron motion.
 The frequency of ZB is $\omega\approx E_g/\hbar$, where $E_g$ is the
energy gap. The amplitude of ZB
 is determined by the strength of periodic potential and the lattice
period and it can be of the
 order of nanometers.  We show that the amplitude of ZB does not
depend much on the width of the wave
 packet representing an electron in real space.
 An analogue of the Foldy-Wouthuysen transformation, known from
relativistic quantum
 mechanics, is introduced in order to decouple electron states in
various bands. We demonstrate that,
 after the bands are decoupled, electrons should be treated as
particles of a finite size.
 In contrast to nearly-free electrons we consider a two-band model of
tightly bound electrons.
 We show that also in this case the electrons should experience the
trembling motion. It is concluded
 that the phenomenon of Zitterbewegung of electrons in crystalline
solids is a rule rather than an exception.
 \end{abstract}

submitto{\SST}
\maketitle

\section{Introduction}
Zitterbewegung (the trembling motion) was theoretically devised by
Schroedinger \cite{Schroedinger30}
after Dirac had proposed his equation describing free relativistic
electrons in vacuum.
Schroedinger showed that, due to a non-commutativity of the quantum
velocity
$\hat{\bm v}=\partial \hat{H}_D/\partial {\bm p}$ with the Dirac
Hamiltonian $\hat{H}_D$,
relativistic electrons experience the Zitterbewegung (ZB) even in
absence of external fields.
The frequency of ZB is about $\omega = 2m_0c^2/\hbar$ and its
amplitude is about the
Compton wavelength $\lambda_c=\hbar/m_0c \approx 3.86\times 10^{-
3}$\AA.
It was later understood that the phenomenon of ZB is due to an
interference of electron states
with positive electron energies ($E>m_0c^2$) and those with negative
energies ($E<m_0c^2$),
see \cite{RoseBook,BjorkenBook,GreinerBook}.
In other words, the ZB results from the structure of the Dirac
Hamiltonian, which contains
both positive and negative electron energies, and it is a purely
quantum effect as it goes
beyond Newton's first law.

An important step in the understanding of ZB was made by Foldy and
Wouthuysen \cite{Foldy54},
(see also \cite{Pryce48,Tani51}), who showed that in absence of
external fields
there exists a unitary transformation that
transforms the Dirac Hamiltonian into a Hamiltonian  in which
positive and negative electron
energies are decoupled. While solutions of the Dirac equation are
four-component functions,
the transformed states for the positive energies have only two upper
non-vanishing
components and those for the negative energies have only two
lower non-vanishing components. Now the above mentioned interference
between
the positive and negative energy states can not occur and there is no
ZB.
Instead, in the new representation the electron is not a point-like
particle, but it
acquires a 'quantum radius' of the size $\lambda_c$. The
interpretation of the two pictures
is until present not quite clear, see
\cite{Newton49,Huang52,Feshbach58,Lock79,Barut68,Barut81,Krekora06}.
To our knowledge, the ZB for free electrons has
never been directly observed. However,  in the presence of the
Coulomb potential
the ZB is manifested in appearance of the so called Darwin term
\cite{RoseBook,BjorkenBook,GreinerBook}.

It was pointed out some time ago that the Zitterbewegung also may
occur in non-relativistic two-band systems in solids
\cite{Cannata90}. It was shown that, similarly to the relativistic
case in vacuum discussed above, the consequence of the ZB is that it
is impossible to localize the electron better than to a certain
finite volume. Recently, an analogy between the Dirac description of
electrons in vacuum and the coupled-band ${\bm k}\cdot{\bm p}$
formalism for electrons in narrow-gap semiconductors (NGS) and
carbon nanotubes (CNT) was used to demonstrate that the ZB should
occur in these systems \cite{Zawadzki05KP,Zawadzki05Nano}. It was
shown that, in agreement with the 'semi-relativistic' analogy
\cite{ZawadzkiOPS,ZawadzkiHMF}, the ZB frequency is always $\omega
\approx E_g/\hbar$, where $E_g$ is the energy gap between the
conduction and valence bands. The amplitude of Zitterbewegung in NGS
and CNT was estimated to be $\lambda_Z=\hbar/m_0^*u$, where $m_0^*$
is the effective electron mass and $u\approx 10^8$cm/s is the
maximum electron velocity in the system. The ZB length in  NGS and
CNT turns out be $10-100$\AA, i.e. $10^4- 10^5$ times larger than in
vacuum. A much lower ZB frequency and its much higher amplitude, as
compared to vacuum, should make the ZB much more readily observable
in semiconductors. The Zitterbewegung was also recently proposed in
two-dimensional systems exhibiting  spin splitting due to structure
and bulk inversion asymmetry \cite{Schliemann2005}, and in 2D
graphite \cite{Katsnelson05}. A phenomenon similar to the ZB was
proposed for electrons in degenerate valence bands in the presence
of an external electric field \cite{Jiang2004}. Very recently, a
unified description of the Zitterbewegung of electrons in different
solid state systems was attempted \cite{Cserti06}.

In view of this recently published work we want to investigate the
question of whether
the phenomenon of Zitterbewegung in solids is a rule rather than an
exception or {\it vice versa}.
To this end we consider two limiting models for electrons in solids:
nearly-free electrons,
for which the periodic potential of the lattice may be treated as a
week perturbation,
and tightly-bound electrons, for which the periodic potential may not
be treated as a
perturbation. Since we are interested in general properties of the
Zitterbewegung,
we do not insist on details of the band models in question but rather
concentrate on
essential features that result in this phenomenon. Although we deal
with non-relativistic
electrons in solids, we use methods of relativistic quantum mechanics
to investigate
an alternative picture in which the trembling motion is replaced by a
kind of
electron 'smearing' in real space. The reason, that a somewhat
mysterious phenomenon of
Zitterbewegung of electrons in vacuum has never been observed, seems
to be related to
its very high frequency and very small amplitude.
The corresponding phenomenon in solids would have much
lower frequency and much larger amplitude. The underlying hope
motivating our work is,
that a more thorough theoretical understanding of the trembling
motion will lead to an
experimental detection of the phenomenon. This would not only deepen
our knowledge of electrons
in solids but also represent a great success of the relativistic
quantum theory.

Our paper is organized in the following way. In Section II we give
the basis of nearly-free electron formalism, Section III treats
the resulting Zitterbewegung using Schroedinger's method of the
equation of motion. In Section IV a more realistic description of
the ZB is presented in which electrons are treated as wave
packets. In Section V we use the Foldy-Wouthuysen transformation
known from the relativistic quantum mechanics to obtain an
alternative electron picture. Section VI treats the Zitterbewegung
in case of tightly bound electrons. In Section VII we discuss the
obtained results and confront them with the previous work. The
paper is concluded by a summary.

\section{Nearly-free electrons}
The beginning of this Section is standard, but it is needed for
further developments. We consider an electron in the presence of
an external periodic potential $V({\bm r})=V({\bm r} + {\bm r_a})$,
where
${\bm r_a}$ is a translation vector of the lattice. The periodic
potential $V({\bm r})$ may be expressed by the Fourier series
$V({\bm r})=\sum_{{\bm l}}V_{{\bm l}}\exp(i{\bm l}{\bm r})$, where
${\bm
l}$ are reciprocal lattice vectors and $V_{{\bm l}}$ are Fourier
components of the periodic potential. For a real potential there
is $V_{{\bm l}}^*=V_{-{\bm l}}$. The wave function of an electron
has the Bloch form
\begin{equation} \label{PsiBloch}
\Psi_{{\bm k}}({\bm r}) = \frac{1}{\sqrt{\cal V}}e^{i{\bm
k}{\bm r}}\sum_{{\bm l}}a_{{\bm l}}e^{i{\bm l}{\bm r}},
\end{equation}
where ${\cal V}$ is the crystal volume and ${\bm k}$ is the wave
vector. Inserting the wave function $\Psi_{{\bm k}}({\bm r})$ into the
Schroedinger equation one obtains the well-known equation for the
energy $E$ and the coefficients $a_{{\bm l}}$
\begin{equation} \label{EqPertb}
 \left( E - \frac{\hbar^2}{2m_0}({\bm k}+{\bm l})^2 \right) a_{{\bm
l}} = \sum_{{\bm g}}V_{{\bm g}}a_{{\bm l}-{\bm g}},
\end{equation}
where $m_0$ is the free electron mass and ${\bm g}$ are reciprocal
lattice vectors.

In absence of the periodic potential there is $a_{{\bm l}}=0$,
$a_0=1$, and $E=\hbar^2{\bm k}^2/2m_0\equiv \epsilon_{{\bm k}}$ is
the free
electron energy. For a weak periodic potential we may treat
$V({\bm r})$ as a perturbation and approximate $a_{{\bm l}}$ for ${\bm
l}\neq 0$ by retaining only linear terms in $V_{{\bm l}}$. We
obtain then
\begin{equation} \label{aPertb}
 a_{{\bm l}} = \frac{V_{{\bm l}}}{\epsilon_{{\bm k}}-\epsilon_{{\bm
k}+{\bm l}}}
\end{equation}
and  $a_0=1$. The perturbed energy is
\begin{equation} \label{EPertb}
 E = \epsilon_{{\bm k}} + \sum_{{\bm l}\neq 0}\frac{|V_{{\bm
l}}|^2}{\epsilon_{{\bm k}}-\epsilon_{{\bm k}+{\bm l}}}.
\end{equation}
For weak potentials the correction to the free electron energy is
small. This, however, is true only if $\epsilon_{{\bm k}} \neq
\epsilon_{{\bm k}+{\bm l}}$. For
${\bm k}$  and ${\bm l}={\bm q}$ such that $\epsilon_{{\bm k}}
=\epsilon_{{\bm k}+{\bm q}}$ we
expect $a_{{\bm q}}$ to be comparable to $a_0$ and the potential
may not be treated as a weak perturbation. The well known way to
treat this problem is to use the approximation for nearly
degenerate levels in which we neglect in (\ref{EqPertb}) all
$a_{{\bm l}}$ except $a_0$ and $a_{{\bm q}}$. We then find
\begin{equation} \label{Eq2x2} \left\{
  \begin{array}{ccc} (E-\epsilon_{{\bm k}})a_0&=& V_{-{\bm q}}a_{{\bm
q}} \\  (E-\epsilon_{{\bm k}+{\bm q}})a_{{\bm q}} &=& V_{{\bm q}}
a_0.  \\\end{array}
 \right. \end{equation}

Equations (\ref{Eq2x2}) are equivalent to dealing with the Hamiltonian
 \begin{equation} \label{defhH} \hat{H} =
   \left(\begin{array}{cc} \epsilon_{{\bm k}+{\bm q}}  & V_q \\
V_q^* & \epsilon_{{\bm k}} \\ \end{array}\right), \end{equation}
which is valid for ${\bm k}$ such that $\epsilon_{{\bm k}+{\bm q}}
\approx \epsilon_{{\bm k}}$.
Hamiltonian (\ref{defhH}) has two eigen-energies
 \begin{equation} \label{defE12}
  E_{1(2)} = \frac{\epsilon_{{\bm k}+{\bm q}}+\epsilon_{{\bm k}}}{2}
\pm \sqrt{|V_{{\bm q}}|^2 + \left(\frac{\epsilon_{{\bm k}+{\bm q}}-
\epsilon_{{\bm k}}}{2}\right)^2}, \end{equation}
and two eigen-states
\numparts
\begin{eqnarray}
   |1> = \frac{1}{N} \left( \begin{array}{c} {E_{\Delta}} + \Delta \\
 V_{{\bm q}}^* \\\end{array} \right) \\
    |2> = \frac{1}{N} \left( \begin{array}{c}  -V_{{\bm q}} \\
{E_{\Delta}} + \Delta \\\end{array} \right),
\end{eqnarray}
\endnumparts

where
\begin{equation} \label{defd}
 \Delta=\frac{1}{2}(\epsilon_{{\bm k}+{\bm q}} - \epsilon_{{\bm k}})
= \frac{\hbar^2}{2m_0}\left( {\bm k} {\bm q} +\frac{{\bm
q}^2}{2}\right), \end{equation}
\begin{equation} \label{defEd} {E_{\Delta}}= \sqrt{|V_{{\bm
q}}|^2+\Delta^2}  \end{equation}
and $N=\sqrt{2{E_{\Delta}}({E_{\Delta}}+\Delta)}$.

\begin{figure}
\includegraphics[width=8.5cm,height=8.5cm]{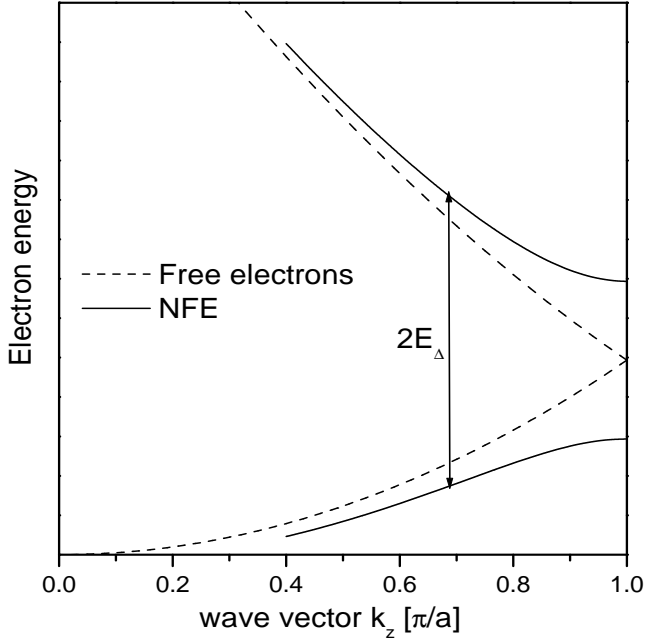}
\caption{ \label{Figure1} Energy {\it versus} $k_z$ for free and nearly-
free electrons
    (schematically). The dashed line shows $E=\hbar^2k_z^2/2m_0$
dispersion  for free electrons
    plotted in the first Brillouin zone. The solid line indicates
$E(k_z)$ dispersion for the
    free motion weakly perturbed by a periodic potential.}
\end{figure}

Taking only one vector ${\bm q}$ in the reciprocal lattice we are
in reality considering a one-dimensional problem. To fix our
attention we consider the symmetry of a simple cubic lattice. As
to the two points in the reciprocal lattice, see (\ref{Eq2x2}),
it is convenient to take ${\bm l}=0$ and ${\bm
l}={\bm q}=[0,0,-2\pi/a]$, where $a$ is the lattice period. Then
the two parabolas $\epsilon_{{\bm k}}$ and $\epsilon_{{\bm k}+{\bm
q}}$ in (\ref{EPertb}) cross
at ${\bm k}=[0,0,\pi/a]$, which determines the Brillouin zone
boundary on the positive $k_z$ axis, see figure \ref{Figure1}. The
energy gap at the zone boundary is $E_g=2|V_{{\bm q}}|$. The
energies $\Delta$ and ${E_{\Delta}}$  depend in reality only on
$k_z$. When
using the nearly-degenerate perturbation theory based on
(\ref{defhH}) we should keep in mind that this procedure is only
valid near the degeneracy point $k_z=\pi/a$, but it progressively
ceases to work as $k_z$ is lowered toward zero.

We note that ${E_{\Delta}}$ of (\ref{defEd}) is analogous to the
relativistic dispersion relation $E(p)
=[(m_0c^2)^2+c^2p^2]^{1/2}$. Thus $|V_{{\bm q}}|^2=(E_g/2)^2$
corresponds to $(m_0c^2)^2$, while $\Delta^2$, which is quadratic
in momentum, corresponds to $c^2p^2$.

It is convenient to split the  Hamiltonian (\ref{defhH}) into  two
parts
\begin{equation} \label{decomposition}  \hat{H} = \hat{H}_{\Delta} +
\hat{H_{\bm k}}, \end{equation}
where
 \begin{equation} \label{defHd} \hat{H}_{\Delta} =
\left(\begin{array}{cc} \Delta  & V_{{\bm q}} \\  V_{{\bm q}}^{*} & -
\Delta \\
  \end{array}\right) \end{equation}
and
 \begin{equation} \label{defHk} \hat{H_{\bm k}} =
\frac{1}{2}(\epsilon_{{\bm k}+{\bm q}} +\epsilon_{{\bm k}})
\left(\begin{array}{cc} 1 & 0\\ 0 & 1 \\
  \end{array}\right). \end{equation}
In terms of the Pauli matrices the Hamiltonian (\ref{defHd}) reads
\begin{equation}  \hat{H}_{\Delta}= \Delta\hat{\sigma}_z +
 \Re(V_{{\bm q}})\hat{\sigma}_x -\Im(V_{{\bm q}})\hat{\sigma}_y.
 \end{equation}

Hamiltonian $\hat{H}_{\Delta}$ has the form reminiscent of the Dirac
Hamiltonian for relativistic electrons in vacuum, while the part
$\hat{H_{\bm k}}$ is proportional to the unity matrix and it can
be treated as a $c$-number. The decomposition
(\ref{decomposition}) is directly related to the two terms in the
energy (\ref{defE12}).

The quantum velocity is $\hat{\bm v}=\partial\hat{H}/\partial(\hbar{\bm k})$.
We calculate
\begin{equation}  \label{vQuant}
  \hat{\bm v}= \hat{\bm v}_{\Delta} + \hat{\bm v}_{{\bm k}} =
           {\bm u}_{\Delta} \hat{\sigma}_z  + {\bm u}_{\bm k}
           {\bm 1},
\end{equation}
where ${\bm u}_{\Delta}=\partial\Delta/\partial(\hbar{\bm
k})=(\hbar/m_0){\bm q}/2$ and ${\bm u}_{{\bm k}}=(\hbar/m_0)({\bm
k}+{\bm q}/2)$. It follows from (\ref{vQuant}) that the quantum
velocity $ \hat{\bm v}$ is an operator, not a number. Since
$\Delta$ depends only on $k_z$, the only non-vanishing component
of $\hat{\bm v}_{\Delta}$ is $\hat{v}_{\Delta z}$. In the
following we drop the index $z$.

 Eigen-values of the quantum velocity  $ \hat{v}_{\Delta}$ are $\mp
(\hbar/m_0)(\pi/a)$. This seems
 paradoxical, as it means that the quantum velocity takes only two
constant (and extreme) values.
 A similar result is obtained for the Dirac equation describing
relativistic
 electrons in vacuum, for which the eigenvalues of the quantum
velocity are $\pm c$.
 It is known that this feature is related to the phenomenon of
Zitterbewegung.

\section{Zitterbewegung}
It can be easily verified that the quantum velocity (\ref{vQuant})
does not commute with the Hamiltonian (\ref{defhH}). This means that
$d\hat{\bm v}/dt$ does not vanish. The original derivation of
Schroedinger's is based on the quantum equation of motion (see also
\cite{Barut81}). Let us calculate the time dependence of $\hat{\bm
v}$. We have
\begin{equation} i\hbar\frac{d\hat{\bm v}}{dt} = [\hat{v}_{\Delta},
\hat{H}] + [\hat{\bm v}_{\bm k}, \hat{H}]. \end{equation}
Since $\hat{H_{\bm k}}$ and $\hat{\bm v}_{\bm k}$ are unity
matrices, they commute with any number matrices. Therefore
$d\hat{\bm v}_{{\bm k}}/dt =1/(i\hbar)[\hat{\bm v}_{{\bm
k}},\hat{H}]=0$, so that $\hat{\bm v}_{{\bm k}}(t)={\bm v}_{{\bm
k} 0}$. Thus
\begin{equation} \label{vprim} i\hbar\frac{d\hat{v}_{\Delta}}{dt} =
[\hat{v}_{\Delta}, \hat{H}_{\Delta}] = 2
\hat{v}_{\Delta}\hat{H}_{\Delta} -
\{\hat{v}_{\Delta},\hat{H}_{\Delta}\}, \end{equation}
where the anti-commutator of
$\{\hat{v}_{\Delta},\hat{H}_{\Delta}\}=2{u_{\Delta}}\Delta$. Hence
\begin{equation}\label{vprim2}  i\hbar\frac{d\hat{v}_{\Delta}}{dt} =
2 \hat{v}_{\Delta}\hat{H}_{\Delta} -2{u_{\Delta}}\Delta. \end{equation}

Let us calculate the second time derivative of $\hat{v}_{\Delta}$
\begin{eqnarray} i\hbar\frac{d^2\hat{v}_{\Delta}}{dt^2}
   &=&  [\frac{d\hat{v}_{\Delta}}{dt}, \hat{H}_{\Delta}] =
\frac{1}{i\hbar}[2\hat{v}_{\Delta}\hat{H}_{\Delta} -
2{u_{\Delta}}\Delta,\hat{H}_{\Delta}]= \nonumber \\
   &=&
\frac{2}{i\hbar}[\hat{v}_{\Delta},\hat{H}_{\Delta}]\hat{H}_{\Delta} =
2\frac{d\hat{v}_{\Delta}}{dt}\hat{H}_{\Delta}. \end{eqnarray}
This represents a differential equation for $d\hat{v}_{\Delta}/dt$.
Its solution is
\begin{equation} \frac{d\hat{v}_{\Delta}}{dt} = \hat{A}_0\exp(-
2i\hat{H}_{\Delta} t/\hbar) \label{vprim_t}, \end{equation}
where $\hat{A}_0$ is a constant operator. Inserting
(\ref{vprim_t}) to (\ref{vprim2})
\begin{equation} i\hbar \hat{A}_0\exp(-2i\hat{H}_{\Delta} t/\hbar) =
2\hat{v}_{\Delta}\hat{H}_{\Delta} - 2{u_{\Delta}}\Delta
\label{EqA0}. \end{equation}
Solving (\ref{EqA0}) for $\hat{v}_{\Delta}$ we obtain
\begin{equation}  \label{vd_t}
  \hat{v}_{\Delta}(t)= \frac{1}{2}i\hbar \hat{A}_0\exp(-
2i\hat{H}_{\Delta} t/\hbar)\hat{H}_{\Delta}^{-
1}+{u_{\Delta}}\Delta\hat{H}_{\Delta}^{-1}.
 \end{equation}
Integrating (\ref{vd_t}) with respect to time and adding the
term due to $z$ component of $\hat{\bm v}_{{\bm k}}$ we finally
find
\begin{eqnarray} \label{z_t}
        \hat{z}(t)&=& z_0 + v_{k_z0}t  +
{u_{\Delta}}\Delta\hat{H}_{\Delta}^{-1}t + \nonumber \\
           & -& \frac{1}{4}\hbar^2 \hat{A}_0\left(\exp(-
2i\hat{H}_{\Delta} t/\hbar)-1\right)\hat{H}_{\Delta}^{-2}.
\end{eqnarray}
In order to find $\hat{A}_0$ we use (\ref{EqA0}) for $t=0$
\begin{equation}  \hat{A}_0 =  \frac{1}{i\hbar}
\left(2\hat{v}_{\Delta}\hat{H}_{\Delta} - 2{u_{\Delta}}\Delta \right)
=  \label{defA0}
            \frac{2{u_{\Delta}}}{i\hbar} \left(\begin{array}{cc} 0 &
V_{{\bm q}}\\-V_{{\bm q}}^{*}& 0 \\\end{array}\right).
\end{equation}
At $t=0$ there is $\hat{z}(0)=z_0$. Similarly, it follows from
(\ref{vd_t}) and (\ref{defA0}) that
$\hat{v}_{\Delta}(0)$ equals to the $z$ component of the initial
velocity $\hat{\bm v}_{\Delta}$
from (\ref{vQuant}).

In order to interpret the result (\ref{z_t}) we observe that the
eigen-energy of $\hat{H}_{\Delta}$
is $\pm {E_{\Delta}}$, where ${E_{\Delta}}$ is given by
(\ref{defEd}). There is $\hat{H}_{\Delta}^{-1} =
\hat{H}_{\Delta}/{E_{\Delta}}^2$ and
$\hat{H}_{\Delta}^{-2} = 1/{E_{\Delta}}^2$. The exponential term in
(\ref{z_t})  is (see Appendix)

\begin{equation} \label{expH}
\exp\left(\frac{-2i\hat{H}_{\Delta} t}{\hbar}\right) =
\cos\left(\frac{2{E_{\Delta}} t}{\hbar}\right) -
i\frac{\hat{H}_{\Delta}}{{E_{\Delta}}}\sin\left(\frac{2{E_{\Delta}}
t}{\hbar}\right).
\end{equation}

The first three terms in (\ref{z_t}) describe the classical
motion. The last term, according to (\ref{expH}), describes
oscillations with the frequency $\omega = 2{E_{\Delta}}/\hbar$. This
frequency corresponds directly to the interband energy
$2{E_{\Delta}}$, as seen in figure \ref{Figure1}.

 Since the Hamiltonian
$\hat{H}_{\Delta}$ is a matrix, the position $\hat{z}(t)$ is also a
matrix. We have explicitly
\begin{equation} \label{z_tH}
\hat{z}(t) = \left(\begin{array}{cc} z_{11}(t) & z_{12}(t)\\
z_{21}(t)& z_{22}(t) \\ \end{array}\right),\end{equation} where
\begin{equation} \label{z11_t}
 \hat{z}_{11}(t) = \frac{\hbar {u_{\Delta}} |V_{{\bm
q}}|^2}{2{E_{\Delta}}^3}\sin\left(\frac{2{E_{\Delta}}
t}{\hbar}\right) +
   \frac{{u_{\Delta}} \Delta^2 t}{{E_{\Delta}}^2} + v_{k_z0}t + z_0.
\end{equation}
The component $\hat{z}_{22}(t)$ has negative signs of the first two
terms. Further
\begin{eqnarray}
 \hat{z}_{21}(t)&=&-\frac{\hbar{u_{\Delta}} V_{{\bm
q}}^*}{2{E_{\Delta}}^2}\left\{
                    i\left[\cos\left(\frac{2{E_{\Delta}}
t}{\hbar}\right) - 1\right] + \right.
        \nonumber  \\ \label{z21_t}
               && \left. +  \frac{\Delta}{{E_{\Delta}}}
\sin\left(\frac{2{E_{\Delta}} t}{\hbar}\right)  \right\}
                 + \frac{{u_{\Delta}} \Delta V_{{\bm q}}^*
t}{{E_{\Delta}}^2},
\end{eqnarray}
where $z_0=z(0)$ and $ \hat{z}_{12}(t) = \hat{z}_{21}(t)^*$.

The amplitude of the oscillating term in (\ref{z11_t}) is
$\hbar{u_{\Delta}}|V_{{\bm q}}|^2/2{E_{\Delta}}^3\approx
\pi\hbar^2/(2m_0a|V_{{\bm q}}|) = \lambda_Z/2$, where the
Zitterbewegung length is defined as
\begin{equation} \label{deflambdaZ}
\lambda_Z = \frac{\pi\hbar^2}{m_0a|V_{{\bm q}}|}.
\end{equation}
It corresponds to the Compton wavelength in relativistic quantum
mechanics. Its numerical estimation is given below.

In agreement with the history of the subject, as described in the
Introduction, we can legitimately call
the above oscillations the Zitterbewegung. We shall discuss the
subject of ZB more thoroughly below.
Here we emphasize how {\it little} we have assumed to obtain the
trembling motion - we have only
perturbed the free electron motion by a periodic potential.

\section{Wave packet}
Now we consider a more realistic picture describing an electron in
terms of a wave packet. Taking, as before, ${\bm q} =
[0,0,-2\pi/a]$ and using the fact that $\Delta$ and ${E_{\Delta}}$
depend
only on $k_z$, we take a gaussian packet in the form
\begin{eqnarray}
 \psi(z) = \frac{1}{\sqrt{2\pi}}\frac{d^{1/2}} {\pi^{1/4}}
           \int_{\infty}^{\infty}\exp\left(-\frac{1}{2}d^2(k_z-
k_{z0})^2\right) \times \nonumber \\
            \exp(i{k_zz}) dk_z   \left( \begin{array}{c}  1 \\  0
\\\end{array} \right), \label{defPacket}
\end{eqnarray}
where $d$ determines the packet's width and $k_{z0}$ fixes its center in
the $k_z$  space. Averaging the oscillating part of the motion over
the wave packet
we obtain from (\ref{z_tH})
\begin{eqnarray} \label{pack_z11_t}
\lefteqn{<\psi(z)| \hat{z}^{osc}(t)|\psi(z)> =} \nonumber \\
 &&       = \frac{d}{\sqrt{\pi}} \int_{-\infty}^{\infty}
\hat{z}_{11}^{osc}(t)\exp\left(-d^2(k_z-k_{z0})^2\right) dk_z,
\end{eqnarray}
where $\hat{z}_{11}^{osc}(t)$ is given in (\ref{z11_t}).

\begin{figure}
\includegraphics[width=8.5cm,height=8.5cm]{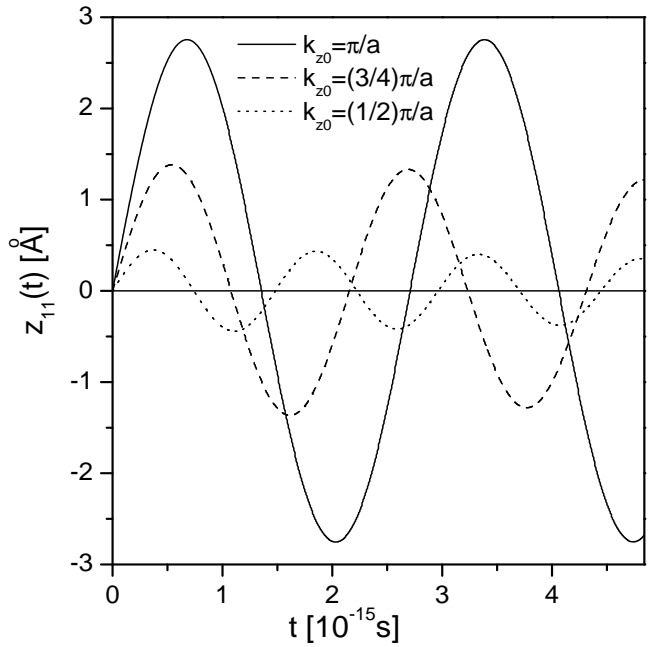}
\caption{\label{Figure2}  Zitterbewegung oscillations of nearly-free
electrons  {\it versus} time,
          calculated for a very narrow wave packet centered  at
various $k_{z0}$ values.
          The band parameters correspond to GaAs, see text.}
\end{figure}

\begin{figure}
\includegraphics[width=8.5cm,height=8.5cm]{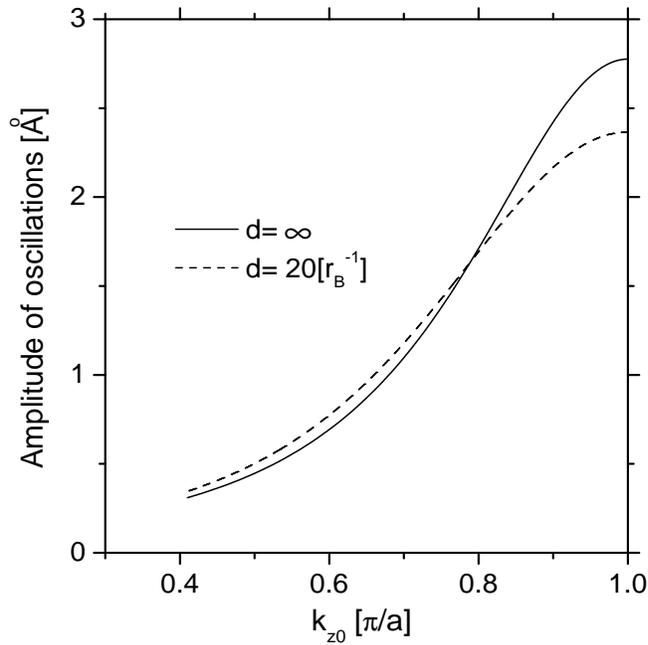}
\caption{\label{Figure3} Amplitude of Zitterbewegung of nearly-free
electrons
        {\it versus} packet center $k_{z0}$, calculated for two
widths of the wave packet.
        Symbol $r_B$ denotes the Bohr radius. Material parameters
correspond to GaAs.}
\end{figure}

To begin, let us take the packet to be a delta function in the
$k_z$ space centered at $k_{z0}$. This corresponds to a completely
non-localized packet in the real space. We can then take various
$k_{z0}$ values beginning with $k_{z0}=\pi/a$ at the zone
boundary. In figure \ref{Figure2} we show the calculated ZB
oscillations of $z_{11}(t)$ for different values of $k_{z0}$. It
can be seen that, as $k_{z0}$ diminishes from $\pi/a$ toward the
zone center, the amplitude of ZB quickly drops. This should not be
surprising since, as is well known (see figure {\ref{Figure1}}), the
effect of the periodic potential on the free electron motion is
the strongest at the zone boundary $k_z=\pi/a$, where the minimum
gap occurs. (We do not consider here the gap at $k_z=0$ for the
upper branch). Figure \ref{Figure3} shows the amplitude of ZB for
$d=\infty$ and $d=20r_B^{-1}$, as calculated from
(\ref{pack_z11_t}). The quantity $r_B=0.53$\AA\ is the Bohr
radius. For $d=\infty$ we deal in  (\ref{defPacket}) and
(\ref{pack_z11_t}) with a delta-function and the solid line in
figure \ref{Figure3} follows the dependence $\hbar{u_{\Delta}}|V_{{\bm
q}}|^2/(2{E_{\Delta}}^3)$ of (\ref{z11_t}). When the width of the
wave
packet increases ($d$ decreases) the amplitude of ZB for
$k_{z0}\approx \pi/a$ diminishes and, as $k_{z0}$ is lowered, it
becomes independent of the width. Since our model is not valid for
$k_{z0}$ near zero, we are limited in our considerations to not
too small $d$ and $k_{z0}$ values.

However, since the ZB amplitude diminishes so quickly with
diminishing $k_{z0}$, it is justified
to limit the considerations of ZB to the vicinity of the band extremes.
It can be seen from figure \ref{Figure2} that, with decreasing $k_{z0}$,
the frequency of ZB
increases. This increase follows from the behavior of the gap
$2{E_{\Delta}}$, as illustrated in
figure \ref{Figure1}.

In order to calculate numerical values of the ZB we need to
specify material parameters. As a matter of example we take
$V_{{\bm q}}=E_g/2=0.76$eV and $a=5.6$\AA, corresponding to GaAs.
This gives $\lambda_Z=5.6$\AA. This value can be compared with
$\lambda_Z=10-13$\AA \ for GaAs, as obtained with the use of ${\bm
k}\cdot{\bm p}$ theory for the fundamental gap in GaAs at ${\bm k}=0$
\cite{Zawadzki05KP}. The above estimation of $\lambda_Z$
based on the simple model is better than one could expect.
Clearly, if we take the $V_{{\bm q}}$ value corresponding to
$E_g=0.23$eV for InSb, $\lambda_Z$ would be seven times larger.

\begin{figure}
\includegraphics[width=8.5cm,height=8.5cm]{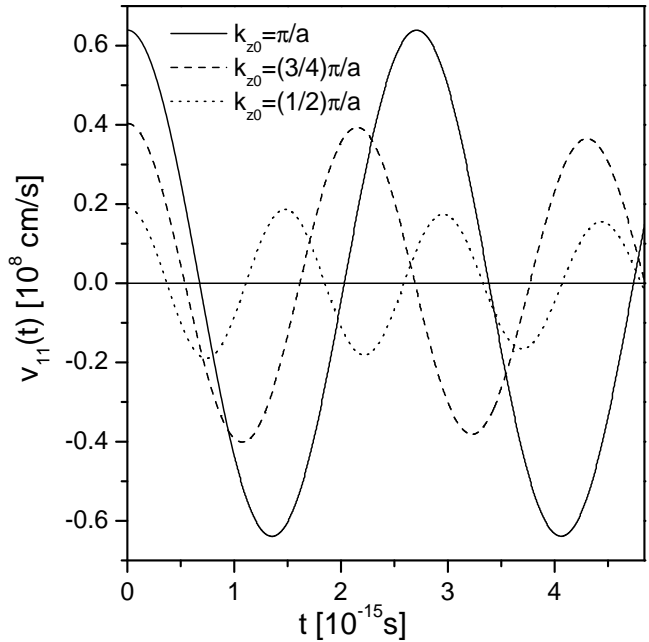}
\caption{ \label{Figure4}
Zitterbewegung contribution to the velocity for nearly-free
electrons  {\it versus} time, calculated for a very narrow wave packet centered  at
various $k_{z0}$ values. The band parameters correspond to GaAs, see text.}
\end{figure}

Next we calculate an observable quantity that is the electric
current caused by ZB. It is given by the velocity multiplied by
the charge. The oscillatory part of velocity is given by the first
term in (\ref{vd_t}). We average it using the wave function
(\ref{defPacket}) which selects the component $\hat{v}_{11}(t)$
\begin{equation} \label{pack_v11_t}
\hat{v}_{11}(t) = \frac{u_{\Delta} |V_{\bm q}|^2}{E_{\Delta}^2}\cos\left(\frac{2E_{\Delta}t}{h}\right)
       + \frac{u_{\Delta}\Delta^2}{E_{\Delta}^2}.
\end{equation}
The results for the velocity, computed with the gaussian wave packet (\ref{defPacket}),
are plotted in figure \ref{Figure4}.
They are quantitatively similar to those shown in figure \ref{Figure2}, the ZB
frequency is clearly the same, the amplitude decreases with increasing $k_{z0}$.
The phase, however, is different and the velocity $\hat{v}_{11}(t)$ is not zero at $t=0$,
but is equal to $u_{\Delta}$ from (\ref{vQuant}). Also, the velocity
does not oscillate around zero, it oscillates around the value $v^{(0)}_{11}$ resulting
from the second term in (\ref{pack_v11_t}). For $k_{z0}=\pi/a$ we can find $v^{(0)}_{11}$ analytically
\begin{equation}
v^{(0)}_{11} = u_{\Delta}  \{1  - \sqrt{\pi} \zeta \exp(\zeta^2) [1-{\rm erf}(\zeta)] \},
\end{equation}
where $\zeta = d/\lambda_Z$ and ${\rm erf}(x)$ is the error function.
For narrow packets (large $d$) the shift $v^{(0)}_{11}$ tends to zero.

\section{Foldy-Wouthuysen transformation}

As mentioned in the Introduction, Foldy and Wouthuysen
\cite{Foldy54} proposed a transformation which, in absence of
external fields, transforms the Dirac Hamiltonian for relativistic
electrons in vacuum into a form in which positive and negative
electron energies are separated. It was recently shown by Zawadzki
\cite{Zawadzki05KP,Zawadzki05Nano} that similar transformations
exist for the ${\bm k} \cdot {\bm p}$ Hamiltonians describing band
structures in narrow gap semiconductors and carbon nanotubes.
Since also the Hamiltonian $\hat{H}_{\Delta}$ of  (\ref{defHd})
bears similarity to the Dirac Hamiltonian, we can expect that a similar
transformation exists for nearly-free electrons as well. This is
indeed the case.

We define a unitary transformation
\begin{equation} \label{defU}  \hat{U} =  \frac{{E_{\Delta}} +
\hat{\beta}
\hat{H}_{\Delta}}{\sqrt{2{E_{\Delta}}({E_{\Delta}}+\Delta)}},
\end{equation}
where $\hat{\beta}=\left(\begin{array}{cc} 1 & 0 \\0  & -1
\\\end{array}\right)$. It is easy to verify that
$\hat{U}\hat{U}^{\dag}=1$. Further
\begin{equation}  \hat{U}\hat{H}_{\Delta} \hat{U}^{\dag} =
{E_{\Delta}}\hat{\beta}  \end{equation}
and obviously $\hat{U}\hat{H_{\bm k}} \hat{U}^{\dag} = \hat{H_{\bm
k}}$ since $\hat{H_{\bm k}}$ is proportional to the unity matrix
(see (\ref{defHk})). Thus, for the transformed Hamiltonian the
eigen-energy problem factorizes into two independent problems for
positive and negative ${E_{\Delta}}$ energies. This means that the wave
function corresponding to the positive energy has the lower
component equal to zero while the wave function for the negative
energy has the upper component equal to zero. However, this is
true also for other wave functions in the transformed
representation, as we show below.

We consider an arbitrary wave function $\Psi(z)$ in the two-component
representation. It can
be expressed in general in the form
\begin{equation}
 \Psi(z) =\int_{-\infty}^{\infty} u(k_z^{'})\exp(ik_z^{'}z)dk_z^{'} = \Psi_+(z) +  \Psi_-(z),
\end{equation}
where
\begin{equation}
 \Psi_{\pm}(z) =\frac{1}{2}\int_{-\infty}^{\infty}  \left(1 \pm
\frac{\hat{H}_{\Delta}}{{E_{\Delta}}}
\right)u(k_z^{'})\exp(ik_z^{'}z)dk_z^{'}.
\end{equation}
Let us now transform the above functions using the $\hat{U}$ operator:
$\Psi^{'}_{\pm}(z)=\hat{U}\Psi_{\pm}(z)$. After some manipulations we obtain
\begin{eqnarray}
\Psi_{\pm}^{'}(z^{'})&=& \frac{1\pm\hat{\beta}}{2} \int_{-
\infty}^{\infty}
\left(\frac{{E_{\Delta}}}{2({E_{\Delta}}+\Delta)}\right)^{1/2} \times
  \nonumber \\
 && \times\left(1 \pm \frac{\hat{H}_{\Delta}}{{E_{\Delta}}}
\right)u(k_z^{'})\exp(ik_z^{'}z^{'})dk_z^{'}.
\end{eqnarray}
The $k^{'}_z$-dependent function under the square root may be put under the integral sign, as
explained in \cite{Foldy54}.

The functions $\Psi_{\pm}^{'}(z^{'})$ have the above mentioned
property of having only the upper or the lower non-vanishing
components, which is guaranteed by the pre-factors
$(1\pm\hat{\beta})$. Using the inverse Fourier transform
\begin{equation}
u(k_z^{'}) = \frac{1}{2\pi}\int_{-\infty}^{\infty} \Psi(z^{'})\exp(-
ik_z^{'}z^{'})dz^{'}, \end{equation}
we have
\begin{equation}
\Psi_{\pm}^{'}(z) = \int_{-\infty}^{\infty} {\cal K}^{\pm}(z,z^{'})\Psi(z^{'})dz^{'},
\label{defPsiPMp} \end{equation}
where
\begin{eqnarray}  {\cal K}^{\pm}(z,z^{'}) =  \frac{1\pm\hat{\beta}}{2} \frac{1}{2\pi}
   \int_{-\infty}^{\infty}
\left(\frac{{E_{\Delta}}}{2({E_{\Delta}}+\Delta)}\right)^{1/2} \times
   \nonumber \\
   \left(1 \pm \frac{\hat{H}_{\Delta}}{{E_{\Delta}}}
\right)\exp(ik_z^{'}(z-z^{'}))dk_z^{'}.    \label{defKPM}
\end{eqnarray}
The kernels ${\cal K}^{\pm}(z,z^{'})$  are not point transformations.
To illustrate this we will transform the eigen-function of the
position operator $\hat{z}^{'}$ in the
old representation, i.e. the Dirac delta function multiplied by a
unit vector
\begin{equation} \Psi(z^{'}) = \delta(z^{'}-z_0)
\left(\begin{array}{c} 1 \\0 \\\end{array}\right). \end{equation}
The transformed functions are
\begin{equation} \Psi^{'}_+(z) = {\cal K}^+_{11}(z,z_0) \label{Phi+}
     \left(\begin{array}{c} 1 \\0 \\\end{array}\right) \end{equation}
and
\begin{equation} \Psi^{'}_-(z) = {\cal K}^-_{21}(z,z_0)
   \left(\begin{array}{c} 0 \\1 \\\end{array}\right),\end{equation}
where
\begin{equation} \label{defKp11}
{\cal K}^+_{11}(z,z_0) = \frac{1}{2\pi\sqrt{2}}\int_{-
\infty}^{\infty} \sqrt{1+\frac{\Delta}{{E_{\Delta}}}}
e^{ik^{'}_z(z-z_0)} dk^{'}_z
\end{equation}
and
\begin{equation}
{\cal K}^-_{21}(z,z_0) = \frac{-V_{{\bm q}}}{2\pi\sqrt{2}}\int_{-
\infty}^{\infty}
   \frac{
e^{ik^{'}_z(z-z_0)}}{\sqrt{{E_{\Delta}}({E_{\Delta}}+\Delta)}} dk^{'}_z.
\end{equation}
It is seen explicitly that the transformed functions
$\Psi^{'}_+(z)$ and $\Psi^{'}_-(z)$ have
vanishing lower or upper components, respectively.
Both ${\cal K}^+_{11}(z,z_0)$ and  ${\cal K}^-_{21}(z,z_0)$ are
normalized to delta functions.
To demonstrate that the transformed functions are characterized by a
certain width, we
calculate their second moments ${\cal M}_2^{\pm}$. Since ${\cal
K}^+_{11}(z,z_0)={\cal K}^+_{11}(\zeta)$,
where $\zeta=z-z_0$, we have
\begin{eqnarray}
{\cal M}_2^+&=& \int_{-\infty}^{\infty}  ({\cal
K}^+_{11}(\zeta))^{\dag}\zeta^2 {\cal K}^+_{11}(\zeta)d\zeta =
\nonumber \\
&=& \frac{1}{(2\pi)^2}\int\!\!\int\!\!\int\! e^{-ik^{'}_z\zeta}
   \sqrt{1+\frac{\Delta_{k^{'}_z}}{E_{\Delta_{k^{'}_z}}}} \times
\nonumber \\
&&  e^{ik^{''}_z\zeta}
\sqrt{1+\frac{\Delta_{k^{''}_z}}{E_{\Delta_{k^{''}_z}}}}
 \zeta^2d\zeta dk^{'}_z dk^{''}_z\!.  \label{defM2}
\end{eqnarray}
In (\ref{defM2}) we used subscripts $k^{'}_z$ and $k^{''}_z$ to indicate the variables of the
integration. Using relations $\zeta =
i(d/dk^{'}_z)\exp(-ik^{'}_z\zeta)$ and similarly for $k^{''}_z$
the triple integral is reduced to
\begin{equation}
{\cal M}_2^+ = \frac{1}{2\pi} \int_{-\infty}^{\infty}
\left(\frac{d}{dk^{'}_z}
    \sqrt{1+\frac{\Delta_{k^{'}_z}}{E_{\Delta_{k^{'}_z}}}} \
\right)^2 dk^{'}_z
 = \frac{1}{32}\lambda_Z \label{WynM2},
\end{equation}
where $\lambda_Z$ is defined in (\ref{deflambdaZ}). The above
result
has been calculated directly using $\Delta$ and ${E_{\Delta}}$ from
(\ref{defd}) and (\ref{defEd}).
Note that since ${\cal K}^+_{11}(\zeta)$ has the dimension [m$^{-1}$],
the dimension of ${\cal M}_2^+$ is [m].

Similar calculations for the second moment of ${\cal K}^-
_{21}(\zeta)$ give
${\cal M}_2^-=\lambda_Z/32$. Thus the transformed functions for the
upper and
lower energies are characterized by the same widths, as should be
expected. We will discuss physical
implications of the above calculations in Section VII.

When transforming various wave functions from the two-component
representation to the one-component
representation with the use of (\ref{defPsiPMp}), it is important
to know more about the
kernels ${\cal K}^{\pm}(z-z^{'})$, as given by (\ref{defKPM}). As an example we will calculate
and plot ${\cal K}^+_{11}(\zeta)$ given by (\ref{defKp11})

Because both $\Delta$ and ${E_{\Delta}}$ are centered at $k^{'}_z=-q_z/2=(\pi/a)$,
it is convenient to change the variables $k^{'}_z \rightarrow k_z - q_z/2$. Then we obtain
\begin{equation} \label{defSmooth} {\cal K}^+_{11}(\zeta) = \exp(-
iq_z\zeta/2)\times K^+_{11}(\zeta), \end{equation}
where
\begin{equation} \label{defK11}
K^+_{11}(\zeta) =  \frac{1}{2\pi\sqrt{2}}\int_{-\infty}^{\infty}
\sqrt{1 + \frac{\Delta^0}{{E_{\Delta}}^0}}\exp(ik_z\zeta) dk_z,
\end{equation}
\begin{equation}  \Delta^0 = \frac{\hbar^2}{2m_0}k_zq_z, \end{equation}
and ${E_{\Delta}}^0=\sqrt{|V_{{\bm q}}|^2+(\Delta^0)^2}$. The
quantities
$\Delta^{0}$ and ${E_{\Delta}}^0$ are centered at $k_z=0$. After the
change
of variables, we singled out the rapidly oscillating part of
${\cal K}^+_{11}(\zeta)$, which is related to the position of
bands extremes in the ${\bm k}$ space. The remaining part
$K^+_{11}(\zeta)$ is a smoothly varying function of $\zeta$ with a
singularity at $\zeta=0$.

\begin{figure}
\includegraphics[width=8.5cm,height=8.5cm]{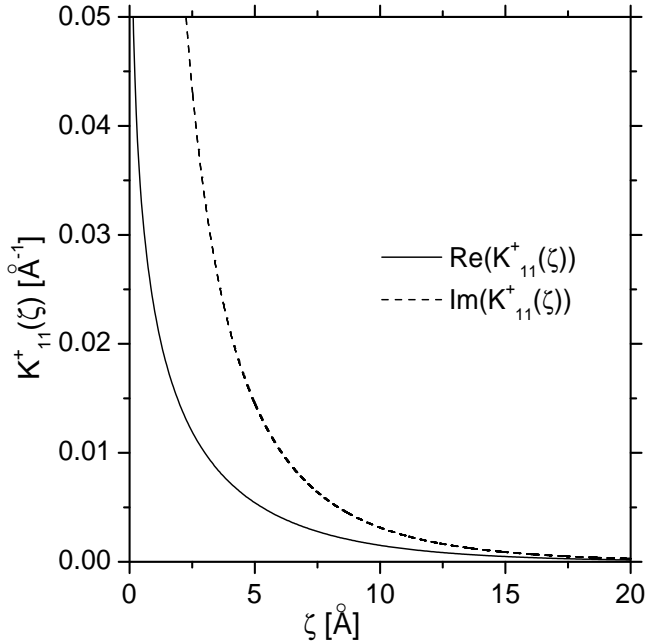}
\caption{ \label{Figure5} Real and imaginary parts of the kernel
of Foldy-Wouthuysen
        transformation for nearly-free electrons (the smooth
component) {\it versus}
        $\zeta=z-z_0$. Material parameters correspond to GaAs.}
\end{figure}

We consider the integrand in (\ref{defK11})
\begin{equation} B(k_z) = \frac{1}{\sqrt{2}} \sqrt{1 +
\frac{\Delta^0}{{E_{\Delta}}^0}}. \label{defB}
\end{equation}
For $k_z\rightarrow \infty$ the function  $B(k_z)$
tends to unity as $1-O(k_z^2)$, while for $k_z\rightarrow -\infty$ it
tends to zero as $O(k_z)$.
Therefore the integral (\ref{defB}) is poorly convergent.
Nevertheless, we can calculate it
with the help of the Heaviside function $\Theta(k_z)$, which has a
similar behavior to the integrand
(\ref{defB}) for $k_z\rightarrow\pm\infty$. We have
\begin{eqnarray} K^+_{11}(\zeta) = \frac{1}{2\pi}&&
   \left( \int_{-\infty}^{\infty} \left[ B(k_z)-
\Theta(k_z)\right]\exp(ik_z\zeta)dk_z + \right.  \nonumber \\
   &&+\left. \int_{-\infty}^{\infty} \Theta(k_z)\exp(ik_z\zeta)dk_z
\right).
\end{eqnarray}
The second integral is  $\int_{-\infty}^{\infty}
\Theta(k_z)\exp(ik_z\zeta)dk_z = i/\zeta + \pi\delta(\zeta)$.
Thus  the real and imaginary parts of $K^+_{11}(\zeta)$ are
\begin{equation} \label{KzRe} \Re[K^+_{11}(\zeta)] = \frac{1}{2\pi}
    \int_{-\infty}^{\infty} \left[ B(k_z)-
\Theta(k_z)\right]\cos(k_z\zeta)dk_z + \frac{1}{2}\delta(\zeta)
\end{equation}
and
\begin{equation} \label{KzIm} \Im[K^+_{11}(\zeta)] = \frac{1}{2\pi}
    \int_{-\infty}^{\infty} \left[ B(k_z)-
\Theta(k_z)\right]\sin(k_z\zeta)dk_z + \frac{1}{2\pi\zeta}.
\end{equation}
The above integrals are carried out numerically. The results are
plotted in figure \ref{Figure5}.
Both $\Re[K^+_{11}(\zeta)]$ and $\Im[K^+_{11}(\zeta)]$ are singular
at $\zeta=0$ and for
large $\zeta$  they decay exponentially.

Since we are dealing with weakly convergent integrals it is of
interest to verify the accuracy of
the above numerical calculations. To this end, we check two sum rules
holding for
$K^+_{11}(\zeta)$ defined in (\ref{defK11}) (see \cite{RoseBook}).
The first rule is
\begin{equation}  \label{defS0}
{\cal S}^+_0=\int_{-\infty}^{\infty} K^+_{11}(\zeta)d\zeta =
\left.  \frac{1}{\sqrt2}  \sqrt{1+\frac{\Delta^0}{{E_{\Delta}}^0}}\
\right|_{k_z=0} =  \frac{1}{\sqrt{2}}.
\end{equation}
Now we verify that the same result is obtained from the numerical
calculations of integrals in  (\ref{KzRe}) and (\ref{KzIm}).
Since the imaginary part of $K^+_{11}(\zeta)$
is an odd function of $\zeta$, it gives no contribution to ${\cal
S}_0$.
The numerical calculation of the real part of $K^+_{11}(\zeta)$
gives  $1/\sqrt{2}$ with the accuracy of 10$^{-3}$.

The second sum rule is related to the second moment of
$K^+_{11}(\zeta)$.
Calculations similar to the  presented above give
\begin{eqnarray} \label{defS2}
{\cal S}^+_2 &=&\int_{-\infty}^{\infty} K^+_{11}(\zeta)\zeta^2d\zeta
=   \nonumber  \\
 &=& \left. -\frac{1}{\sqrt2}
\frac{d^2}{dk^2_z}\sqrt{1+\frac{\Delta^0}{{E_{\Delta}}^0}}\
\right|_{k_z=0}  =
\frac{1}{\sqrt2}\left(\frac{\lambda_Z}{2} \right)^2.
\end{eqnarray}
Taking into account the normalization (\ref{defS0}) we obtain the
extension of $K^+_{11}$ function to be $\lambda_Z/2$, which is in
exact analogy to
the relativistic Dirac electrons \cite{RoseBook}, see also
\cite{Zawadzki05KP}.
Similar results are obtained for the ${\cal S}^-_0$ and ${\cal S}^-
_2$ integrals defined
using $K^-_{21}$ function.

The FW transformation is not the only transformation which can decouple positive and negative
energies in the field-free case. In the relativistic quantum mechanics described by the Dirac equation
other transformations were devised, see for example Cini-Touschek  \cite{Cini58}. In a recent paper
Mulligan \cite{Mulligan06} introduced a still different transformation separating the $4\times 4$
Dirac equation into two $2\times 2$ equations for the electron and antielectron, respectively.
It is possible that an analogous transformation would be possible for the nearly-free non-relativistic
electrons considered above.

\section{Tightly bound electrons}
In the preceding sections we considered the case of a weak periodic
potential acting on free electrons and
we showed that this potential leads to the ZB. Now we consider an
opposite limit of strong
periodic potential.

An effective treatment of a strong periodic potential is the
tight-binding method. We use here as an example the so called
Empirical Tight Binding Method. In this model one takes one $s$
orbital per cation  and three $p$ orbitals per anion including
nearest-neighbor and second nearest-neighbor interactions.
Spherical approximation is assumed, so all $sp$ and $p$ bands are
isotropic and their ${\bm k}$-dependence is given by the
$\Gamma-X$ dispersion. The model was used for calculating magnetic
interactions in dilute magnetic semiconductors, approximating the
band structure of Cd$_{1-x}$Mn$_x$Te within the  whole Brillouin
zone \cite{Larson88}. This scheme provides a good
semi-quantitative description of both upper valence bands as well
as the lowest conduction band. In the basis $x_a,y_a,z_a,s_c$,
(c-cation, a-anion) the Hamiltonian is
 \begin{equation} \label{defHETMB}
 \hat{H} = \left(\begin{array}{cccc}
     t^{(2)}_k & 0         & 0         & iV_k  \\
     0         & t^{(3)}_k & 0         & 0     \\
     0         & 0         & t^{(3)}_k & 0     \\
     -iV_k     & 0         & 0         & t^{(1)}_k
    \end{array}\right), \end{equation}
where
 \begin{equation} t^{(1)}_k = \epsilon_c+ 4C[1+2\cos(\frac{1}{2}ak)],
\label{defT1}  \end{equation}
 \begin{equation} t^{(2)}_k = \epsilon_a+
4A_2+8A_1\cos(\frac{1}{2}ak)],  \end{equation}
 \begin{equation} t^{(3)}_k = \epsilon_a+ 4A_1[1+\cos(\frac{1}{2}ak)]
+ 4A_2\cos(\frac{1}{2}ak), \label{defT3}  \end{equation}
 \begin{equation} V_k = 4V_{ca}\sin(\frac{1}{4}ak).  \label{defVk}
\end{equation}
Six parameters in (\ref{defT1})-(\ref{defVk}):
$\epsilon_c=3.16$eV, $\epsilon_a=0.1$eV, $V_{ca}=1.103$eV,
$C=0.015$eV, $A_1=0.13$eV, $A_2=0.15$eV  are the
Slater-Koster parameters in notation used in  \cite{Larson88},
and $a=6.482$\AA\ is CdTe lattice constant.
The parameters $\epsilon_c$ ($\epsilon_a$)
are cation (anion) on-site energies, $V_{ca}$ is a single nearest-
neighbor hopping parameter,
while $C, A_1, A_2$ are the second-neighbor parameters.

Hamiltonian (\ref{defHETMB}) can be factorized giving one doubly
degenerate energy band  $E_{2,3}= t_3(k)$
coming from $p_y$ and $p_z$ orbitals, and two energy bands coming
from the $s-p_x$ interaction.
The $sp_x$ Hamiltonian can be written in the form
 \begin{equation} \label{Hsp}
 \hat{H}_{sp} =  \left(\begin{array}{cc}\Delta_k &-iV_k \\iV_k& -
\Delta_k \end{array}\right)  +
                 \Gamma_k \left(\begin{array}{cc}1 &0\\0&1
\end{array}\right),
\end{equation}
where $\Delta_k=\frac{1}{2}(t^{(1)}_k-t^{(2)}_k)$
  and $\Gamma_k=\frac{1}{2}(t^{(1)}_k+t^{(2)}_k)$.
The Hamiltonian (\ref{Hsp})  is very similar to that for
nearly-free electrons of (\ref{decomposition})-(\ref{defHk}).
The difference is, that $\Delta_k$ and $V_k$ depend on the
absolute value of $|{\bm k}|$. Also, now the minimum band gap
occurs at $k=0$. All eigen-energies of the Hamiltonian
$\hat{H}_{sp}$ have periodicity of the lattice constant. The
quantum velocity $\hat{\bm v}=\partial\hat{H}_{sp}/(\partial{\hbar
\bm k})$ does not commute with $\hat{H}_{sp}$. To calculate
$\hat{r}(t)$ we use  the Heisenberg picture
\begin{equation}
 \hat{r}(t) = \exp(i\hat{H}_{sp}t/\hbar)\hat{r}(0)\exp(-
i\hat{H}_{sp}t/\hbar),
\end{equation}
which gives  for the $\hat{r}(t)$ matrix
\begin{eqnarray}
\hat{r}_{11}(t)&=&
\left(\frac{V_k\Delta_k^{'}V_k-V_k\Delta_k V_k^{'}}{2{E_{\Delta}}^3}\right)
             \sin\left(\frac{2{E_{\Delta}} t}{\hbar}\right) +
 \nonumber \\ \label{z11_etbm}
  &+& \left(\frac{\Delta_k V_k
V_k^{'} +\Delta_k^2\Delta_k^{'}}{{E_{\Delta}}^2\hbar}\right)t +
v_{\Gamma}t + r_0.
\end{eqnarray}
The component $\hat{r}_{22}(t)$ is given by (\ref{z11_etbm}) with
the changed signs of the first two terms. Further
\begin{eqnarray} \label{z21_etbm}
\hat{r}_{21}(t)&=& \frac{\Delta_k
V_k^{'}-\Delta_k^{'}V_k}{2{E_{\Delta}}^2} \nonumber
 \left \{ \left[\cos\left(\frac{2{E_{\Delta}} t}{\hbar}\right)-
1\right] - i\frac{\Delta}{{E_{\Delta}}}\times  \right.  \\
  &&\left. \sin\left(\frac{2{E_{\Delta}} t}{\hbar}\right) \right\}
  -i\left(\frac{V_k^2V_k^{'} + V_k\Delta_k\Delta_k^{'}}{\hbar
{E_{\Delta}}^2}\right)t,
\end{eqnarray}
and $r_{12}=r_{21}^*$.
In  ($\ref{z11_etbm}$)-($\ref{z21_etbm}$) the prime denotes
a differentiation with respect to $k$,
and $v_{\Gamma}=\partial \Gamma_k/(\partial \hbar k)$.
Equations ($\ref{z11_etbm}$)-($\ref{z21_etbm}$) are formally similar
to (\ref{z11_t})-(\ref{z21_t}),
the main difference is that now there appear terms related to the
dependence of $V_k$ on $k$.

\begin{figure}
\includegraphics[width=8.5cm,height=8.5cm]{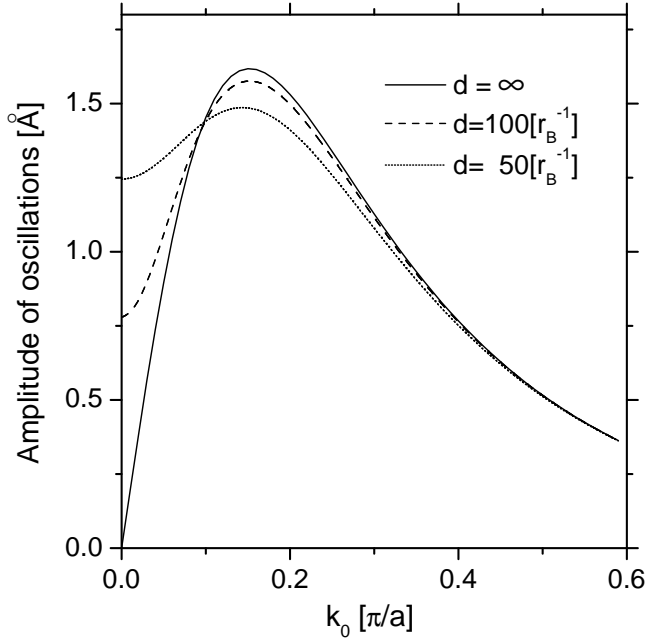}
\caption{ \label{Figure6} Amplitude of Zitterbewegung of tightly
bound electrons
        {\it versus} packet center $k_0$, calculated for three widths
of the wave packet.
        Symbol $r_B$ denotes the Bohr radius. Material parameters
correspond to CdTe, see text.}
\end{figure}

We calculated the ZB  oscillations of $\hat{r}_{11}(t)$, as given
by (\ref{z11_etbm}), using the gaussian packet of the form
(\ref{defPacket}). The results are similar to those illustrated in
figure \ref{Figure2}. In figure \ref{Figure6} we show the
calculated amplitudes of ZB as functions of the packet center
$k_0$ for three widths of the packet. Here the $k_0$-dependence
does not have a maximum at $k_0=0$, because at this point the
interaction $V_k$ between the bands vanishes, see (\ref{defVk}).

The presence of a strong periodic potential leads to two effects.
First, the quadratic dispersion relation $E_{{\bm k}}\propto {\bm
k}^2$ for free electrons is replaced by a periodic one. Second,
the potential mixes $s$ and $p_x$ states to form two $sp_x$ energy
bands. This mixing leads to the Zitterbewegung.

\section{Discussion}
The main result of our work is that both in the case of nearly-free
electrons and in the opposite
case of tightly bound electrons we predict the Zitterbewegung
phenomenon. Comparing
this result with the previous work, in which the ZB was predicted
with the use of LCAO \cite{Cannata90}
and ${\bm k}\cdot{\bm p}$ theory \cite{Zawadzki05KP,Zawadzki05Nano},
we conclude
that ZB is not due to a particular approach. In fact, the mathematics
is quite
similar in all above theories and, although we deal with non-
relativistic  electrons, it resembles
the formulation of relativistic quantum mechanics for free electrons
in vacuum. It is clear that
the fundamental underlying reason for the appearance of ZB in solids
is the periodic potential of
the lattice. Particularly instructive in this respect are figures
\ref{Figure2} and \ref{Figure3} of the present paper which show that the amplitude of
ZB is directly related to the effect of periodic potential on the free electron
motion.

This result is not surprising. Without specifying any particular band
model we deal in solids
with the periodic Hamiltonian $\hat{H} = \hat{\bm p}^2/2m_0 + V({\bm
r})$ and the electron velocity is
$\hat{\bm v}=\partial\hat{H}/\partial\hat{\bm p} = \hat{\bm p}/m_0$.
It follows that
$d\hat{\bm v}/dt=1/(i\hbar)[\hat{\bm v},\hat{H}]\neq 0$, i.e. the
velocity is not constant in time
because of the periodic potential $V({\bm r})$. In this perspective
the various models mentioned above
simply illustrate how this result comes about.

Clearly, the velocity does not commute with the Hamiltonian in the
presence of other potentials as well. However the periodic
potential is special because, due to the Bloch theorem, the
electrons can propagate in the perfect crystal without scattering
and the quasi-momentum $\hbar{\bm k}$ is a good quantum number.
For this reason it is possible to treat the electrons as almost
free particles and replace the influence of the periodic potential
by an effective electron mass. Still, as demonstrated in Refs.
\cite{Zawadzki05KP,Zawadzki05Nano} and the present paper, within
two-band models the basic non-commutativity  of $\hat{\bm v}$ and
$\hat{H}$ mentioned above remains in the form of non-commuting
$2\times 2$ matrices with the resulting Zitterbewegung. If the
bands  are completely separated, we have $\hat{H}_{eff}=\hat{\bm
p}^2/2m^*$ and $\hat{\bm v}=\hat{\bm p}/m^*$, so that
$\hat{H}_{eff}$ and $\hat{\bm v}$ commute and the ZB disappears.
However, there is a price to pay for this separation. It is shown
in Refs. \cite{Zawadzki05KP,Zawadzki05Nano} and in the present
paper that, once the electrons are described by a one-band
equation (so that their energy is completely specified), they
should be treated as objects of a finite size. The last effect is
observable in the presence of an external potential due to
appearance of the so called Darwin term for free relativistic
electrons \cite{RoseBook,BjorkenBook,GreinerBook,Feshbach58} as
well as semiconductor electrons \cite{Zawadzki05KP}.

As emphasized throughout our paper (see also Refs.
\cite{Zawadzki05KP,Zawadzki05Nano}),
in the two-band model the ZB of non-relativistic
electrons  in solids is in close analogy to the ZB of relativistic
electrons in vacuum,
as first proposed by Schroedinger. The Hamiltonians for the two cases
are very similar
and in both systems the ZB results from an interference of electron
states corresponding to
positive and negative electron energies
\cite{RoseBook,BjorkenBook,GreinerBook}.
If, with the use of Foldy-Wouthuysen transformation, the states of
positive and negative energies
are separated in the Hamiltonian and in the wave functions, the ZB
does not occur because
the positive energy state (say)
has nothing to interfere with. This corresponds to the separation of
bands mentioned above and the
conclusions of the two reasonings agree.

Thus we are confronted with the following choice: 1) We use a two-
band description,
the electrons are point-like particles and  they experience the
Zitterbewegung.
2) We use a one-band description, the electrons do not experience the
Zitterbewegung but they are
   characterized by a quantum radius of the size equal to the ZB
amplitude.

The last point is illustrated by (\ref{defS2}) describing the
average 'smearing' of the transformed delta function.
It is equal to the amplitude of ZB given by (\ref{z11_t}). One
can say that the separation of energy
bands by the Foldy-Wouthuysen transformation is equivalent to a
certain averaging of the ZB motion.

It was observed \cite{Lock79} that, since the Zitterbewegung had been
predicted for plane Dirac waves,
it is not quite clear what the trembling motion means for an electron
uniformly distributed in space.
In this connection it is important that the amplitude does
not vary much when the electron is represented by a wave packet
localized in real space,
see figure \ref{Figure3}.

Passing to more specific points of our treatment, we emphasize
again how little we had to assume to derive the ZB in case of
nearly-free electrons - it was enough to perturb the free electron
motion by a periodic potential. This case has certain
particularities. In the 'typical'  two-band situations, both in
vacuum \cite{RoseBook,GreinerBook,Feshbach58} and in solids
\cite{Zawadzki05KP,Zawadzki05Nano}, there exists a maximum
velocity in the system which plays an important role in the
theory. In case of nearly-free electrons there is no maximum
velocity since the perturbed energy branches tend asymptotically
to the free electron parabola $E=\hbar^2{\bm k}^2/2m_0$, see figure
\ref{Figure1}. This is reflected in the velocity $\hat{\bm v}_{{\bm
k}}$, while the velocity $\hat{v}_{\Delta}$ has the typical
'two-band' behavior
and it is responsible for the ZB. In fact, the velocity
$\hat{v}_{\Delta}$ has
the maximum value. It is equal to $(\hbar/m_0)(\pi/a)$, see our
eigen-value considerations after (\ref{vQuant}). For
$a=5.6$\AA \ the maximum velocity is 6.4$\times 10^7$cm/s, which
should be compared with $u=1.3\times 10^8$cm/s obtained from the
${\bm k}\cdot{\bm p}$ theory for GaAs and other III-V compounds
\cite{Zawadzki05KP}. Again, our simple model gives quite a
reasonable estimation. In the nearly-free electrons model
$\lambda_Z$ is proportional to $1/V_{{\bm q}}\sim 1/E_g$ (see
(\ref{deflambdaZ})), which agrees with the ${\bm k}\cdot{\bm p}$
approach \cite{Zawadzki05KP}, where $\lambda_Z\sim 1/m_0^* \sim
1/E_g$. In narrow gap materials $\lambda_Z$ can be as large as
tens of angstroms.

As far as the phenomenon of ZB is concerned, the behavior of
nearly-free and tightly bound electrons is quite similar. The main
difference comes from the fact that in nearly-free case the
Fourier coefficients of the periodic potential do not depend on
the wave vector ${\bm k}$, whereas in the tightly bound case they
go to zero for vanishing $k$. As a result, in the first case (see
figure \ref{Figure3}) the ZB amplitude is highest for the $k_{z0}$
corresponding to the minimum energy gap, while in the second case
(see figure \ref{Figure6}) the maximum amplitude is shifted with
respect to the minimum gap. For the tightly bound case the
electrons in $t_k^{(3)}$ bands (see (\ref{defT3})) would not
exhibit the ZB, but we do not insist on this point since these two
bands are not realistically described by the model.

In order to illustrate that, after the Foldy-Wouthuysen
transformation is carried out,
the former Dirac delta function is 'smeared' into
the kernel ${\cal K}^+_{11}$ of (\ref{defKp11}), we not only
calculate it numerically (see figure \ref{Figure5}), but also calculate
its second
moment ${\cal M}^+_2$ (see (\ref{WynM2})) and the sum rule ${\cal
S}^+_2$ of (\ref{defS2}).
The second moment has the advantage of using the standard quantum
mechanical probability distribution
$(K^+_{11})^{\dag}K^+_{11}$, see (\ref{defM2}). Its disadvantage
is that this probability distribution
is normalized to the Dirac delta function and not to a number. As to
the sum rules using the single
$K^+_{11}$ function, their advantage is that both the normalization
(\ref{defS0}) and the sum rule
${\cal S}^+_2$  (\ref{defS2}) are numbers, and (\ref{defS2})
can be simply interpreted as a square of 'smearing'
(This procedure was used for free relativistic electrons by Rose
\cite{RoseBook}).
The disadvantage of using the single $K^+_{11}$ is that we have to
separate out its 'smooth' part,
see (\ref{defSmooth}). The oscillatory part $\exp(-iq_z\zeta)$
would not appear
if we considered the nearly-free electron gap at $k_z=0$.

Methods to observe the Zitterbewegung should clearly be adjusted to
the investigated materials, but
it seems that an appropriate tool would be the scanning probe
microscopy  which can produce
images of coherent electron flow \cite{Topinka00,Leroy03}.
This technique uses a sharp mobile tip which can sense the
electron charge. If one used dilute magnetic semiconductors of CdMnTe
type \cite{Larson88},
one could employ magnetic effects caused by electron oscillations.

The second category of possible observable effects is related to the
problem of what happens
when electrons are confined to dimensions smaller than $\lambda_Z$.
In relativistic quantum
mechanics one finds statements that a measurement of the position of
a particle, if carried
out with greater precision than the Compton wavelength, would lead to
pair production \cite{Newton49}.
It is clear, however, that the pairs created this way can only be
virtual, otherwise their
recombination would lead to the production of energy out of nothing.
The virtual
carriers could be observed in screening or in magnetism. An
alternative point of view states
that a stiff confinement is equivalent to an infinite potential well
having the width
$\Delta r < \lambda_Z$, and the electrons will simply occupy the
lowest energy level in such a well.
It is, however, certain that if an electron is confined to dimensions
$\Delta r < \lambda_Z$,
its energy (or its uncertainty) is of the order of the gap $E_g$
between the positive and
negative electron energies, which means that the one-band description
is not adequate.
We are then back to the two-band model, which was our starting point.

There remain many unanswered questions concerning
the trembling motion but it appears that in crystalline solids it
represents a rule rather than
an exception. According to the theory, the ZB in semiconductors has
decisive advantages
over the corresponding effect in vacuum. Thus an experimental
detection of the trembling
motion in solids appears to be a matter of near future.

\section{Summary}
We considered theoretically non-relativistic  nearly-free electrons
in solids for which the
periodic potential of the lattice may be treated as a weak
perturbation on the free electron motion.
Using the two-band model we showed that electrons experience the
trembling motion (Zitterbewegung)
in absence of external fields, similar to that for free relativistic
electrons in vacuum.
The frequency of ZB and its amplitude were derived. The frequency is
$\omega\approx E_g/\hbar$
where $E_g$ is the energy gap between the two bands. The amplitude
$\lambda_Z$
depends on the strength of periodic potential and the lattice period.
For typical parameters $\lambda_Z$ can be of the order of 10\AA\ to
100\AA\ that is $10^4-10^5$
times larger than in vacuum. The trembling motion is also considered
for nearly-free electrons
represented by wave packets, it is shown that the amplitude is not
strongly dependent on packet's
width. The Foldy-Wouthuysen type of unitary transformation, known
from relativistic quantum
mechanics, is used to separate the energy bands. Consequences of the
FW transformation are investigated.
It is demonstrated that, if one uses a one-band description, the
electrons do not experience the
trembling motion but they should be treated as particles having size
$\lambda_Z$. Tightly bound electrons
are considered as well to provide an opposite case to nearly-free
electrons. Within the two-band model
the trembling motion is obtained also in this case demonstrating that
the ZB phenomenon is not related to
a specific theoretical approach. It is concluded that the trembling
motion is directly related to the
effect of the periodic potential on the electron and, as such, it
should occur in many situations in solids.

\ack
This work was supported in part by the Polish Ministry of Sciences,
Grant No PBZ-MIN-008/P03/2003.

\appendix
\section{}
First, we prove some properties of $\hat{H}_{\Delta}$ of
(\ref{defHd}). The
eigen-values of $\hat{H}_{\Delta}$ are $\pm {E_{\Delta}}$. If $|1>$
and $|2>$ are eigen-states of $\hat{H}_{\Delta}$,
then $\hat{P}_i=|i><i|, (i=1,2)$ are two
projection operators. It follows that $\hat{P}_1+\hat{P}_2=1$ and
$\hat{H}_{\Delta}={E_{\Delta}}(\hat{P}_1-\hat{P}_2)$. Accordingly
\begin{equation} (\hat{P}_1-
\hat{P}_2)=\frac{\hat{H}_{\Delta}}{{E_{\Delta}}}. \end{equation}
Since $\hat{H}_{\Delta}^{-1}=(1/{E_{\Delta}})(\hat{P}_1-\hat{P}_2)$,
we have  $\hat{H}_{\Delta}^{-1}=\hat{H}_{\Delta}/{E_{\Delta}}^2$.
Then $\hat{H}_{\Delta}^{-2} = \hat{H}_{\Delta}^{-1}\hat{H}_{\Delta}^{-
1} = 1/{E_{\Delta}}^2$ because $\hat{H}_{\Delta}^2={E_{\Delta}}^2$.
For a real $b$ there is
\begin{equation} \exp(i\hat{H}_{\Delta} b) = \exp(i{E_{\Delta}}
b)\hat{P}_1 + \exp(-i{E_{\Delta}} b)\hat{P}_2. \end{equation}
Because $\exp(\pm i{E_{\Delta}} b) = \cos({E_{\Delta}} b) \pm
i\sin({E_{\Delta}} b)$ we have
\begin{eqnarray}
\exp(i\hat{H}_{\Delta} b)&=& [\cos({E_{\Delta}} b) +
i\sin({E_{\Delta}} b)]\hat{P}_1 + \nonumber \\
            &+& [\cos({E_{\Delta}} b) - i\sin({E_{\Delta}}
b)]\hat{P}_2.
\end{eqnarray}
Grouping the terms with cosine and sine functions we get
\begin{eqnarray}
\exp(i\hat{H}_{\Delta} b)\!\!&=&\!\cos({E_{\Delta}}
b)(\hat{P}_1\!+\!\hat{P}_2) +   i\sin({E_{\Delta}} b)(\hat{P}_1\!-
\!\hat{P}_2)  =  \nonumber \\
            \!\!&=&\!\cos({E_{\Delta}} b) + i
\frac{\hat{H}_{\Delta}}{{E_{\Delta}}} \sin({E_{\Delta}} b).
\end{eqnarray}
This identity is used in (\ref{expH}).

\end{document}